\documentclass[11pt]{article}
\usepackage{graphics}   % use the graphics package
\usepackage{amssymb}
\usepackage{amsmath}
\usepackage{amsthm}
\newtheorem{theorem}{Theorem}
\newtheorem{corollary}{Corollary}

\newtheorem{lemma}{Lemma}
\newtheorem{fact}{Fact}

\title{On Precision - Redundancy Relation in the Design of Source Coding Algorithms}

\author{
Yuriy A. Reznik
\\
Qualcomm Incorporated \\
5775 Morehouse Drive, San Diego, CA 92122 \\
E-mail: {\tt yreznik@ieee.org \/} }

\date{}

\begin{document}

\maketitle

\begin{abstract}
We study the effects of finite-precision representation of source's
probabilities on the efficiency of classic source coding algorithms,
such as Shannon, Gilbert-Moore, or arithmetic codes. In particular,
we establish the following simple connection between the redundancy
$R$ and the number of bits $W$ necessary for representation of
source's probabilities in computer's memory ($R$~is assumed to be
small):
\begin{equation*}
W \lesssim \eta \log_2 \frac{m}{R} \,,
\end{equation*}
where $m$ is the cardinality of the source's alphabet, and $\eta
\leqslant 1$ is an implemen\-ta\-tion-specific constant. In case of
binary alphabets ($m=2$) we show that there exist codes for which
$\eta = 1/2$, and in $m$-ary case ($m > 2$) we show that there exist
codes for which $\eta = m/(m+1)$. In general case, however (which
includes designs relying on progressive updates of frequency
counters), we show that $\eta = 1$. Usefulness of these results for
practical designs of source coding algorithms is also discussed.
\end{abstract}

\section{Introduction}
Since Shannon it is known that the average rate of a code
constructed for a stochastic source cannot be lower than the entropy
of this source. The difference between the rate and the entropy is
called the {\em redundancy\/} of the code.

It is also known, that the redundancy of a code fundamentally
depends on the number of input symbols that are jointly mapped into
a code during the encoding process. This number (say, $n$) is
usually called a {\em block-size\/} or {\em delay\/} of the code.
The ratio of redundancy over $n$ is called the {\em redundancy
rate\/} of a code, and the speed of its convergence
%to 0 with increase of $n$
has long been considered a key criterion in understanding the
effectiveness of source codes.

For example, it was shown that many classic codes for known
memoryless sources (such as block Shannon, Huffman, or Gilbert-Moore
codes \cite{CoverThomas,Krichevsky}) attain the redundancy rate of
$R=O(1/n)$~\mbox{\cite{Gallager,Capocelli_deSantis,Krichevsky,Stubley,Szpankowski00}}.
Krichevsky-Trofimov codes for a class of memoryless sources achieve
the rate of $O(\log n/ n)$~\cite{Krichevsky}. Lempel-Ziv codes for
this class were shown to converge at the rate of $O(\log \log n /
\log n)$~\cite{zl0,WynerZiv91}, etc.

%Based on such estimates, one might assume that any given redundancy
%target is achievable by simply taking large enough block size. But,
%in a practical world, such a projected solution may not be feasible
%because of basic hardware constraints, such as finite amount of
%memory, or finite precision of registers that can be used for
%performing arithmetic operations.
%
%Hence, the redundancy of many practically realizable codes depends
%not only on their block-size or delay, but also on resource or
%complexity constraints of their implementations. Popular in practice
%implementations of arithmetic codes
%\cite{Rubin79,Pasco76,WittenNealCleary} represent well recognized
%examples of such complexity - redundancy design tradeoffs.
%

At the same time, much less is known about the connection between
the %achievable
redundancy and typical implementation constraints, such as width of
computer's registers. Perhaps the best example of a prior effort in
exploring this connection was the development of finite-precision
implementations of arithmetic codes (cf. Pasco \cite{Pasco76}, Rubin
\cite{Rubin79}, Witten et al. \cite{WittenNealCleary}). Simple
redundancy bounds for such finite-precision algorithms were offered
by Han and Kobayashi~\cite{HanKobayashi} and Ryabko and
Fionov~\cite{RyabkoFionov}. However, both of these results are
specific to particular implementations.

In this paper, we pose and study somewhat more general questions:
\begin{itemize}
\item {\em "What is the achievable redundancy of a code that can be
constructed by a machine with $W$-bits representations of source's
probabilities?"\/}, and/or
\item {\em "What is the smallest number of bits that one should use for
representing source's probabilities in order to construct a code
with target redundancy $R$?"\/}
\end{itemize}
For~simplicity, we~assume that codes can use infinite delays, and
that the redundancy is caused only by errors of finite precision
representations of probabilities.

Indeed, if the answer to the first question is known, then it can be
treated as a lower bound for achievable redundancy of any source
coding algorithm with such constraint on internal representation of
source's probabilities (or their estimates). The answer to the
second question is of immediate practical interest.

%In our study, we show that this problem can be effectively reduced
%to a precision analysis of rational approximations of a set of real
%numbers. Eventually, the problem becomes a special case of what is
%known in number theory as simultaneous {\em Diophantine
%approximations\/}~\cite{Cassels}, and we use known tools from this
%theory to answer our question.

%Our final result -- formulae connecting redundancy with bit-widths
%and other parameters, reveal a particular (and somewhat surprising)
%sensitivity of this relationship on the size of the alphabet. For
%instance, it turns out that (under some conditions) codes for binary
%sources can be constructed with up to twice-shorter registers than
%equally-redundant codes constructed for sources with large
%alphabets.

This paper is organized as follows. In Section~2, we present formal
setting of our problems and describe our results. Their possible
applications are discussed in Section~3. All proofs~and~supplemental
information from Diophantine appro\-xi\-ma\-tion theory (which we
use in order to derive our results) are given in Appendix~A.

\section{Definitions and Main Results}
Consider an $m$-ary memoryless source $S$, producing symbols
$\alpha_1,\ldots,\alpha_m$ with probabilities $p_1,\ldots,p_m$. We
assume that in our source coding algorithm instead of true
probabilities $p_1,\ldots,p_m$, we have to use their approximations:
$\hat{p}_1,\ldots,\hat{p}_m$.

By $\delta_i = p_i - \hat{p}_i$~ ($i=1,\ldots,m$) we denote
probability differences (errors) for each symbol, and by
\begin{equation}
\delta^* = \max_i \left| p_i - \hat{p}_i \right| = \max_i \left|
\delta_i \right| \label{eq:delta_star}\,,
\end{equation}
we denote their maximum absolute value. Further, by $p_{\min}$
we denote the smallest probability in the source:
\begin{equation}
p_{\min} = \min_i p_i
\end{equation}
and we will assume that $p_{\min} > 0$, and that it is also
relatively large compared to our maximum approximation error:
\begin{equation}
\frac{\delta^*}{p_{\min}} < 1\,.
\end{equation}

\subsection{Upper Bounds for Redundancy}
We first show that, under above described conditions, the loss in
compression efficiency incurred by using
$\hat{p}_1,\ldots,\hat{p}_m$ is bounded by the following simple
expression.
\begin{lemma}
\begin{equation}
D\left(p\,||\,\hat{p}\right) \leqslant m \delta^* \frac{1}{1 -
\delta^* / p_{\min}} = m \delta^* \left(1 +
O\left(\delta^*\right)\right)\,. \label{eq:D_via_delta}
\end{equation}
\end{lemma}

We next consider a more specific source coding scheme, in which
source's probabilities are approximated by rational values:
\begin{equation}
~~~~~~~\hat{p}_i = \frac{f_i}{t}\,, ~~~f_i,t \in
\mathbb{N}~~~~(i=1,\ldots,m) \label{eq:p_hat_frac}
\end{equation}
where
\begin{equation}
t = \sum_{i=1}^m f_i\,
\end{equation}
is the common denominator.

We derive two results regarding attainable redundancy due to such
approximations. We start with a more general (and weaker) statement:
\begin{theorem}
Given any $t \gg 1$, it is possible to find rational approximations
of source's probabilities (\ref{eq:p_hat_frac}), such that:
\begin{equation}
D\left({p\,||\,f/t}\right) \, \leqslant \, \frac{m}{2\,t} \,
\frac{1}{1 - 1 / \left({2\,t\, p_{\min}}\right)} = \frac{m}{2\,t} \,
\left( 1 + O\left(t^{-1}\right)\right)\,. \label{eq:D_via_t}
\end{equation}
\end{theorem}
This shows that redundancy is decreasing approximately inverse
proportionally to $t$. This bound is correct for any values $t$.

Nevertheless, a more detailed study of this problem (which involves
the use of tools and results from Diophantine approximation
theory~\cite{Cassels}) reveals that among various possible values of
denominator $t$, there exists some, for which precision of
approximation (\ref{eq:p_hat_frac}) can be much higher. In turn,
this leads to much lower redundancy of source codes based on such
approximations.

We claim the following:
\begin{theorem}
There exist infinitely many integers $t, f_1,\ldots,f_m$ $(m>2)$
producing approximations of source's probabilities
(\ref{eq:p_hat_frac}), such that:
\begin{equation}
D\left({p\,||\,f/t}\right) \, < \, \frac{m}{t^{1+1/m}} \, \frac{1}{1
- 1 / \left({t^{1+1/m}\, p_{\min}}\right)} = \frac{m}{t^{1+1/m}}
\,\left(1 + O\left(t^{-1-1/m} \right)\right)\,.
\label{eq:D_via_t_mary}
\end{equation}

In a case when ~$m=2$, there exist infinitely many integers $t, f_1$
(with $f_2=t-f_1$) producing approximations of source's
probabilities (\ref{eq:p_hat_frac}), such that:
\begin{equation}
D\left({p\,||\,f/t}\right) \, < \,
\frac{2\, \kappa}{t^2} \frac{1}{1 - \kappa / \left( t^2\, p_{\min} \right)} = %
\frac{2\, \kappa}{t^2} \left(1 +O\left(t^{-2}\right)\right)\,,
\label{eq:D_via_t_binary}
\end{equation}
where:
\begin{equation}
\kappa =
\left[\begin{array}{ll}%
5^{-1/2} \,, & \mbox{if~~} p_1 = \frac{r \psi + s}{u\psi + v},~ \psi
= \frac{\sqrt{5} - 1}{2}\,,~ rv -us
= \pm 1\,,~ r,s,u,v \in \mathbb{Z}\,,\\
2^{-3/2} \,, & \mbox{otherwise\,.} \label{eq:kappa}
\end{array}
\right.
\end{equation}
\end{theorem}

The above result is not immediately obvious, as it implies that
codes relying on rational approximations (\ref{eq:p_hat_frac}) and
constructed for binary sources, can be much more precise than codes
using equally large parameter $t$ but constructed for $m$-ary
sources ($m>2$). In general, based on the above result, the larger
is the cardinality of the alphabet, the more severe is the effect of
approximations of source's probabilities.

Recall, that most traditional redundancy bounds for codes for
memoryless sources (such as delay-redundancy relations, obtained
assuming infinite precision implementation) don't change their order
based on cardinality of the alphabet. Our finding above indicates
that for many practical algorithms this may no longer be a case once
one start accounting redundancy contributions caused by
finite-precision implementation.

\subsection{Precision (and Memory Usage) vs Redundancy}
Consider an implementation of a source coding algorithm employing
rational approximations of source's probabilities
(\ref{eq:p_hat_frac}). We assume that such an algorithm can either
store immediate values $f_i$, or their cumulative equivalents (as it
is convenient, e.g. in the design of Shannon, Gilbert-Moore, or
arithmetic codes):
\begin{equation}
s_i = \sum_{\footnotesize
\begin{array}{c} j=1,\ldots,i \\
\{k_j\}: p_{k_1} \leqslant \ldots \leqslant p_{k_m}
\end{array}} f_{k_j}\,.
\end{equation}
In either case, the maximum integer value that might be
stored per symbol is~$t$. Hence, the required {\em register width\/}
or {\em memory cost per symbol\/} $W$ in such a scheme is simply:
\begin{equation}
W = \lceil {\log_2 t} \rceil.
\end{equation}

Using our prior results, we derive the following simple bounds for
this quantity:

\begin{corollary}
In the design of a source coding scheme, the number of bits $W$ used
for representing source's probabilities, and the resulting increase
in redundancy $R$ satisfy:
\begin{equation}
W < \log_2\left({\frac{m}{R} + \frac{1}{p_{\min}}}\right) =
\log_2\left(\frac{m}{R}\right) + O\left( R \right)
\label{eq:W_general}
\end{equation}
\end{corollary}

\begin{corollary}
There exist infinitely many implementations of source codes, in
which the number of bits $W$ used for representing source's
probabilities and the resulting increase in redundancy $R$ satisfy:
\begin{equation}
W < \frac{m}{m+1}\log_2 \left(\frac{m}{R} + \frac{1}{p_{\min}}
\right) + 1 = \frac{m}{m+1}\log_2 \left(\frac{m}{R}\right) + 1 +
O(R)\,,
\end{equation}
where $m>2$ is a cardinality of source's alphabet.

In a case when ~$m=2$, there exist infinitely many implementations
of binary source codes, in which the number of bits $W$ used for
representing source's probabilities and the resulting increase in
redundancy $R$ satisfy:
\begin{equation}
W < \frac{1}{2} \log_2\left( \frac{2}{R} + \frac{1}{p_{\min}}
\right) + \frac{1}{2} \log_2 4 \kappa  \\
%& = & \frac{1}{2} \log_2\left( \frac{2}{R} \right) + \frac{1}{2}
%\log_2 \kappa + 1 + O\left( R \right) \\
= \frac{1}{2} \log_2\left( \frac{8\,\kappa}{R} \right) + O\left( R
\right)\,,
\end{equation}
where $\kappa$ is a constant defined in (\ref{eq:kappa}).
\end{corollary}

Here, it can be seen that the use of binary alphabets can lead up to
a factor of~2 savings in the number of bits needed in representation
of probabilities.

It should be noted, however, that this factor of 2 does not implies
that conversion to binary alphabets will produce equivalent saving
in overall storage usage. The~problem here is that in order to
implement such codes one need to store not only values $f_1$ or
$s_1$, but also $f_2$ or $s_2$ or $t$.
%
%Hence the total memory usage $M$ (in bits) for such codes is
%\begin{equation}
%M = m W\,,
%\end{equation}
%where $m$ is the cardinality of the alphabet.
%
For example, if one would substitute an $m$-ary source with a
cascade (binary tree) of $m-1$ binary sources, then this would
increase memory usage by a factor of $2(m-1)/m$, prior to a
possibility to use gains predicted by our Corollary 2. Nevertheless,
such a conversion can still be justified by the need to use shorter
registers.

In general, given an $m$-ary source ($m\geqslant2$), we can say that
the amount of memory $M$ that needs to be use for construction
of its code (that involves rational representation of its
probabilities (\ref{eq:p_hat_frac})) is at least
\begin{equation}
M = m W\,,
\end{equation}
where $W$ is connected to redundancy as predicted by our Corollaries
1 and 2.

\section{Conclusions}
Several bounds establishing connection between redundancy and
precision of representation of probabilities in source coding
algorithms have been derived.

These results can be used for maximizing performance of codes with
respect to precision or memory available on each particular
computing platform.

The results of our Theorem 2 (and Corollary 2) revealing the
existence of higher-precision approximations, particularly for
sources with small cardinality of alphabets may also influence
future practical designs of source coding algorithms. For example,
in handling of sources with large alphabets one may consider
grouping of their symbols and using a cascade of codes with smaller
alphabets as a way to improve their performance given register-width
constraints.

\appendix
\section{Proofs}
We first prove statement of our Lemma 1:
\begin{eqnarray}
D\left(p\,||\,\hat{p}\right) & = & \sum_{i=1}^m p_i \log \left(
\frac{p_i}{\hat{p}} \right) \nonumber \\
%& = & \sum_{i=1}^m p_i \log \left( \frac{p_i}{p_i - \delta_i}
%\right) \\
%& = & \sum_{i=1}^m p_i \log \left( \frac{1}{1 - \delta_i / p_i}
%\right) \\
& = &  - \sum_{i=1}^m p_i \log \left( {1 - \delta_i / p_i} \right)
\nonumber \\
& \leqslant & - \sum_{i=1}^m p_i \log \left( {1 - \delta^* / p_i}
\right) \nonumber \\
& \leqslant & \sum_{i=1}^m p_i \frac{\delta^* / p_i}{1 - \delta^* /
p_i} \label{eq:in1} \\
& \leqslant & \sum_{i=1}^m \frac{\delta^* }{1 - \delta^* / p_{\min}}
\nonumber \\
& = & m \delta^* \frac{1 }{1 - \delta^* / p_{\min}} \nonumber
\end{eqnarray}
where (\ref{eq:in1}) is due to the following inequality
\cite{AbramowitzStegun}:
\begin{equation*}
-\log(1 - x) \leqslant \frac{x}{1-x}\, ~~~~~(x < 1).
\end{equation*}

We next consider rational approximations (\ref{eq:p_hat_frac}) of
source probabilities. By assuming that factors $f_i$ are chosen such
that
\begin{equation*}
\delta_i = \left| {p_i - f_i/t} \right| = \frac{1}{t} \left| {t\,p_i
- f_i} \right| = \frac{1}{t}\, \min_{z \in \mathbb{Z}}  \left|
{t\,p_i - z} \right|\,,
\end{equation*}
we can show that:
\begin{equation*}
\delta_i \leqslant \frac{1}{2\,t}\,,
\end{equation*}
and, consequently:
\begin{equation*}
\delta^* \leqslant \frac{1}{2\,t}\,.
\end{equation*}
This, combined with Lemma 1 leads to a statement of our Theorem 1.

In order to prove first statement of Theorem 2, we will need to use
the following result regarding attainable precision of simultaneous
Diophantine approximations \cite[p.\,14, Theorem~III]{Cassels}:
\begin{fact}
For any $n > 1$ irrational values $\alpha_1,\ldots,\alpha_n$, there
exist infinitely many sets of integers $(a_1,\ldots,a_n,q)$ such
that:
\begin{equation*}
q^{1/n} \max\left\{ \left\lvert{q\, \alpha_1 -
a_1}\right\rvert,\ldots, \left\lvert{q\, \alpha_n -
a_n}\right\rvert\right\}< \frac{n}{n+1} \,.
\end{equation*}
\end{fact}
In our case, this means that there must exist rational
approximations of set of probabilities $p_1,\ldots,p_m$, such that:
\begin{equation*}
\delta^* < \frac{m}{m+1} \, t^{-1-1/m} <  t^{-1-1/m}\,.
\end{equation*}
This, combined with Lemma 1, leads to bound (\ref{eq:D_via_t_mary})
claimed by the Theorem 2.

Indeed, it shall be noted, that the task of finding such Diophantine
approximations is not a trivial one, and the reader is referred to a
book of M. Gr\"{o}etschel, L. Lov\'{a}cz, and A. Schrijver
\cite{GroetschelLovaczSchrijver} which discusses this problem in
details.

Consider now binary case ($m=2$). We first notice that since
$p_1+p_2=1$, the problem is essentially reduced to studying
precision of a single approximation:
\begin{equation*}
~~~~~~~~~\hat{p}_1 = f_1/t\,.  ~~~~~~~~(f_1,t \in \mathbb{N})
\label{eq:binary}
\end{equation*}
Thus, by setting $\hat{p}_2 = f_2/t = (t - f_1)/t$ we can see that:
\begin{equation*}
\left|{\delta_2}\right| = \left|{p_2 - f_2/t}\right| = \left|{1 -
p_1 - (t - f_1)/t}\right| = \left|{p_1 - f_1/t}\right| =
\left|{\delta_1}\right| = \delta^*\,.
\end{equation*}

Hence, here we are dealing with a scalar (one-dimensional) case of
Diophantine approximations, and the following result apply
\cite[p.\,11, Theorem~V]{Cassels}:
\begin{fact}
Let $\alpha$ be irrational. Then there are infinitely many $q$ and
$a$ such that
\begin{equation*}
q \left|{q \alpha - a}\right| < 5^{-1/2}.
\end{equation*}
If $\alpha$ is equivalent to $\frac{1}{2}\left(\sqrt{5} - 1\right)$
then the constant $5^{-1/2}$  cannot be replaced by any smaller
constant. If $\alpha$ is not equivalent to
$\frac{1}{2}\left(\sqrt{5} - 1\right)$, then there are infinitely
many $q$ and $a$ such that:
\begin{equation*}
q \left|{q \alpha - a}\right| < 2^{-3/2}.
\end{equation*}
\end{fact}

This means, that in our case with probabilities $p_1,p_2=1-p_1$,
there must exist approximations for which:
\begin{equation*}
\delta^* < \kappa\, t^{-2}\,,
\end{equation*}
where $\kappa$ is a constant depending on $p_1$ in the following way
(which absorbs definition of equivalence implied by above quoted
result):
\begin{equation*}
\kappa =
\left[\begin{array}{ll}%
5^{-1/2} \,, & \mbox{if~~} p_1 = \frac{r \psi + s}{u\psi + v},~ \psi
= \frac{\sqrt{5} - 1}{2}\,,~ rv -us
= \pm 1\,,~ r,s,u,v \in \mathbb{Z}\,,\\
2^{-3/2} \,, & \mbox{otherwise\,.}
\end{array}
\right.
\end{equation*}

By combining this with Lemma 1 we arrive at the second bound
(\ref{eq:D_via_t_binary}) claimed by the Theorem 2.

Our Corollaries 1, and 2 are simple consequences of Theorems 1 and
2, in which we use inequality:
\begin{equation*}
\log_2 t \leqslant W < \log_2 t + 1.
\end{equation*}

\end{document}